\documentclass[aps,prl,superscriptaddress,twocolumn]{revtex4-1}
\usepackage{amsmath}
\usepackage{amssymb}
\usepackage{bm}
\usepackage{epsfig}
\usepackage{graphicx}
\usepackage{color}
\usepackage{ulem}

\begin{document}

\title{Permanent Rabi oscillations in coupled exciton-photon systems with
PT-symmetry}
\author{I.Y. Chestnov}
\affiliation{Department of Physics and Applied Mathematics, Vladimir State University named after A. G. and N. G. Stoletovs, 87 Gorkogo, Vladimir, 600000, Russia}
\author{S.S. Demirchyan}
\affiliation{Department of Physics and Applied Mathematics, Vladimir State University named after A. G. and N. G. Stoletovs, 87 Gorkogo, Vladimir, 600000, Russia}
\author{S.M. Arakelian}
\affiliation{Department of Physics and Applied Mathematics, Vladimir State University named after A. G. and N. G. Stoletovs, 87 Gorkogo, Vladimir, 600000, Russia}
\author{A.P. Alodjants}
\affiliation{Department of Physics and Applied Mathematics, Vladimir State University named after A. G. and N. G. Stoletovs, 87 Gorkogo, Vladimir, 600000, Russia}
\affiliation{Russian Quantum Center, 100 Novaya, Skolkovo, 143025, Moscow Region, Russia}
\author{Y.G. Rubo}
\affiliation{Instituto de Energ\'{\i}as Renovables, Universidad Nacional Aut\'onoma de M\'exico, Temixco, Morelos, 62580, Mexico}
\author{A.V. Kavokin}
\affiliation{Russian Quantum Center, 100 Novaya, Skolkovo, 143025, Moscow Region, Russia}
\affiliation{Physics and Astronomy School, University of Southampton, Highfield,
Southampton, SO171BJ, UK}
\date{\today}

\begin{abstract}
We propose a physical mechanism which enables permanent Rabi oscillations in
driven-dissipative condensates of exciton-polaritons in semiconductor
microcavities subjected to external magnetic fields. The method is based on
incoherent excitonic reservoir engineering. We demonstrate that permanent
non-decaying oscillations may appear due to the parity-time (PT) symmetry of
the coupled exciton-photon system realised in a specific regime of pumping
to the exciton state and depletion of the reservoir. For effective non-zero
exciton-photon detuning, permanent Rabi oscillations occur with unequal
amplitudes of exciton and photon components. Our predictions pave way to
realisation of integrated circuits based on exciton-polariton condensates.
\end{abstract}

\maketitle


\textit{Introduction.---}Nowadays, semiconductor microcavities with embedded
quantum wells serve as a solid-state laboratory for fundamental studies of
dynamic, coherent, nonlinear and quantum effects in nonequilibrium ensembles
of bosonic quasiparticles: exciton-polaritons \cite%
{RMP82.1489.2010,PSS242.2005}. Various approaches to photonic information
processing with the use of microcavity polaritons have been proposed, see,
e.g., \cite{PRB.82.155313.2010,polarit,PRL.112.196403.2014}. In particular,
polariton circuits, or neurons, are proposed in \cite{circuits} and then
experimentally examined by Ballarini et al. in \cite{Nat.com.4.1778.2013} at
extremely low pump intensities. Although nonequilibrium exciton-polariton
Bose-Einstein condensation has been already observed in many labs, the
requirement to use high Q-factor microcavities with high reflectivity Bragg
mirrors limits wide application of such phenomena for practical purposes.
Even in highest quality microcavities used in current experiments the
leakage of photons leads to dissipation effects occurring on the scale of
tens of picoseconds. In particular, this leads to a rapid attenuation of
Rabi oscillations occurring between exciton and photon components of a
polariton state, which represent a particular interest for quantum
information applications. Recently, some of us have
proposed \cite{PRL.112.196403.2014} a way to improve the coherence time of
polariton Rabi oscillations by stimulated pumping of a Rabi-oscillator from
a permanent thermal reservoir of polaritons. The increase of the the
coherence time of polariton Rabi-oscillations in the presence of \textit{cw}
pumping has been experimentally observed \cite{DeGiorgi}, and manifestations
of this effect in spatial dynamics of exciton-polaritons have been revealed
experimentally~\cite{Dominici} and analyzed theoretically \cite{liew14}.
Recently, Voronova \textit{et al} \cite{Voronova} has studied theoretically
the non-linear regime of polariton Rabi-oscillations and their interplay
with exciton-photon Josephson oscillations.

In the Letter, we demonstrate theoretically realisation of permanent Rabi
oscillations between excitonic and photonic components of a
driven-dissipative spinor exciton-polariton condensate in semiconductor
microcavities at specific pumping conditions. We study the regime where the
microcavity system obeys parity-time (PT) symmetry conditions. Originally,
the PT-symmetry approach has been proposed by Bender \textit{et al} \cite%
{bender98,bender99} to demonstrate that non-Hermitian Hamiltonians can
possess entirely real energy eigenvalue spectra. In the past decade,
PT-symmetry features have been demonstrated for various systems in photonics
\cite{ruter10,feng11,szameit11,PRA.82403803.2010,PRA.86.053809.2012,barashenkov12},
condensed matter physics \cite{PRL.110.083604.2013} and metamaterials \cite%
{PRL.110.173901.2013}. Moreover, PT-symmetry approach has been extended to
electronic circuits \cite{Schindler} which makes it highly relevant to
polaritonics. Physically, PT-symmetry requirements can be easily understood
by using the system of two linearly coupled oscillators (dimers, or
waveguides in photonics \cite{PRA.82403803.2010,PRA.86.053809.2012}). The
system obeys PT-symmetry criteria if the rate of dissipation for one of the
oscillators is exactly equal to the rate of gain in the other oscillator.
This condition can be achieved in a system of two coupled polariton
condensates with distributed dissipation rates \cite{aleiner12}, that
results in the conservative Hamiltonian dynamics. It is important to note
that even if the exact compensation between gain and losses in a two-mode
system is not achieved, the concept of quasi-PT-symmetry can be introduced
\cite{j.opt.16.065501.2014}. Here we demonstrate that the regime of
permanent Rabi oscillations can be realized for the exciton-photon system in
a microcavity, and the condition of PT-symmetry may be achieved if the gain
in the excitonic component is compensated by losses from the photonic
component.

\textit{Basic equations.---}We consider a coupled exciton-photon system in
the presence of the external magnetic field that produces the Zeeman
splitting $\hbar \Delta _{Z}$ of exciton levels, and an incoherent excitonic
reservoir, pumped by cw-field $P$. This non-resonant pumping can be realized
both optically or by electronic current injection \cite{Schneider13}. We
assume that the wave function of driven-dissipative exciton-polariton
condensate consists of photonic and excitonic components, $\phi _{\pm }$ and
$\chi _{\pm }$. We will neglect the possible spatial degrees of freedom in
what follows, assuming the condensate to be at zero momentum state. In this
case, both components of the condensate and the reservoir can be described
by Boltzmann kinetic equations:
\begin{subequations}
\label{allsystem}
\begin{eqnarray}
\dot{\chi}_{\pm } &=&\frac{1}{2}\left( p_{X}[N_{\pm }]-\gamma _{X}\right)
\chi _{\pm }+i\delta _{\pm }\chi _{\pm }-i\Omega \phi _{\pm }, \\
\dot{\phi}_{\pm } &=&-\frac{1}{2}\gamma _{P}\phi _{\pm }-i\Omega \chi _{\pm
}, \\
\dot{N}_{\pm } &=&P-\gamma _{R}N_{\pm }-p_{X}[N_{\pm }]\left\vert \chi _{\pm
}\right\vert ^{2},
\end{eqnarray}%
\end{subequations}
where dots denote time derivatives and $2\Omega $ is a polariton Rabi
splitting frequency. The subscript \textquotedblleft $+$\textquotedblright\
(\textquotedblleft $-$\textquotedblright ) corresponds to the spin
projection parallel (antiparallel) to the vector of magnetic field. In Eqs.~%
\eqref{allsystem} $\delta _{\pm }=\Delta \pm \Delta _{Z}$ are effective
photon-exciton detunings determined by the detuning $\Delta =\omega
_{P}-\omega _{X}$ of the cavity mode and exciton frequencies in the absence
of magnetic field and by the Zeeman splitting $\Delta _{Z}$; $\gamma _{X}$, $%
\gamma _{P}$ and $\gamma _{R}$ are the exciton, cavity and reservoir damping
rates, respectively. In Eqs.~\eqref{allsystem} the term containing $%
p_{X}[N_{\pm }]$ describes the pumping of the exciton state by stimulated
scattering from the reservoir. We shall consider two possible scattering
mechanisms described by pumping terms $p_{1}$ and $p_{2}$:
\begin{subequations}
\begin{eqnarray}
p_{X}[N_{\pm }] &\equiv &p_{1}=R_{1}N_{\pm },  \label{RN} \\
p_{X}[N_{\pm }] &\equiv &p_{2}=R_{2}N_{\pm }^{2}\left\vert \chi _{\pm
}\right\vert ^{2}.  \label{RN2Chi2}
\end{eqnarray}
\end{subequations}

The Eq.~\eqref{RN} implies the acoustic phonon assisted pumping with the rate $%
R_{1}$, see \cite{PRL.99.140402.2007}, while Eq.~\eqref{RN2Chi2} describes
the exciton-exciton scattering where excitons possessing momenta $-\mathbf{k}
$ and $\mathbf{k}$ scatter into the condensate state with the momentum $%
\mathbf{k}=0$. Note that in both cases, the scattering feeds both upper and
lower exciton-polariton branches. Our model accounts for the decay of both
exciton and photon components of the condensate but neglects incoherent
processes that lead to relaxation between upper and lower polariton branches.

To start with, let us neglect scattering of excitons with opposite spins so
that Eqs.~\eqref{allsystem} can be solved for each of spin components
separately. We first examine the subsystem with \textquotedblleft $+$%
\textquotedblright\ spin component omitting subscript for simplicity.

The dynamics of the exciton-polariton system demonstrates a threshold
behavior at the specific value $P^{th}$ of the cw-pump which is dependent on
the particular mechanism of the exciton pumping $p_{X}[N]$ \cite%
{PRL.99.140402.2007}. At $P<P^{th}$ the exciton pumping rate $p_{X}[N]$
is comparable to or much smaller than dumping rates $%
\gamma _{X}$ and $\gamma _{P}$ which, in turn, are much smaller than the
Rabi splitting $\Omega $ (see, e.g., \cite{SSC.93.773.1995}). In this case
one can assume the reservoir population $N$ and the pumping rate $p_{X}[N]$ to
be time independent on a time scale of the Rabi
oscillation period $\sim $ $\Omega ^{-1}$. In particular, for the pumping
term given by Eq.~\eqref{RN}, eliminating Eq.~(\ref{allsystem}c) we find
eigenfrequencies of the exciton-photon system
\begin{multline}
\omega _{1,2}=\frac{i}{4}\left( \tilde{\gamma}-p_{X}\right) +\frac{\delta }{2%
}  \label{eigen_without_pump} \\
\pm \frac{1}{2}\sqrt{4\Omega ^{2}+\left( \delta -\frac{i}{2}\left( \gamma
_{P}-\gamma _{X}+p_{X}\right) \right) ^{2}},
\end{multline}%
characterizing steady-state solutions $\chi =\chi _{0}e^{i\omega t}$ and $%
\phi =\phi _{0}e^{i\omega t}$. Physically, Eqs.~%
\eqref{eigen_without_pump} determine frequencies of upper ($\omega _{1}$)
and lower ($\omega _{2}$) polariton branches, cf.~\cite{SSC.93.773.1995}. In
Eq.~\eqref{eigen_without_pump} we denoted $\tilde{\gamma}=\gamma _{P}+\gamma
_{X}$.  In this case, from Eq.~(\ref{allsystem}c) one obtains $P_{1}^{th}=%
\tilde{\gamma}\gamma _{R}/R_{1}$.

The PT-symmetry of the system \eqref{allsystem} appears if the hermitian
part of Hamiltonian is symmetric and the anti-hermitian part is
anti-symmetric with respect to permutation of excitonic and photonic
components. The first condition leads to
\begin{subequations}
\label{PT_condition}
\begin{equation}
\delta =0,
\end{equation}%
and the second condition requires that the losses from the photonic
component should be fully compensated by the gain of the exciton one,
\begin{equation}
p_{X}-\gamma _{X}=\gamma _{P}.
\end{equation}%
Clearly, if these conditions are satisfied the eigenfrequencies $\omega
_{1,2}$ \eqref{eigen_without_pump} become fully real.  We note that in this
case the amplitudes of oscillations of exciton and photon populations become
equal $\left\vert \chi \right\vert ^{2}=\left\vert \phi \right\vert
^{2}=\left\vert A\right\vert ^{2}$. Conditions (\ref{PT_condition}a,b) can
be experimentally realized by tuning of the pumping of the excitonic
component with variation of the \textit{cw} pumping strength.

Figure~\ref{fig_decay} reveals the decay dynamics of a coupled matter-light
system determined by Eqs.~\eqref{allsystem} %
 for different pumping mechanisms. The chosen parameters are relevant to experimentally accessible GaAs
semiconductor microstructures, cf.~\cite{RMP82.1489.2010}. In the Letter for simplicity we represented excitonic $|\chi|^2$ and photonic $|\phi|^2$ populations in the units of initial condensate density. As it is seen
from Fig.~\ref{fig_decay} and follows from Eq.~\eqref{eigen_without_pump},
the characteristic time $\tau _{R}\propto 2/\left( \tilde{\gamma}%
-p_{X}\right) $ of Rabi oscillations decay can be increased by manipulating
the  pump field parameter $p_{X}$.

Condition (\ref{PT_condition}b) is a criterion for realisation of permanent
Rabi oscillations ($\tau _{R}\rightarrow \infty $). It is easy to show that
such a regime cannot be achieved below threshold, at $P<P_{{}}^{th}$, see
Fig.~\ref{fig_decay}. Actually, since pumping rate $p_{X}\equiv p_{X}[N(t)]$
depends on the number of particles $N(t)$ in the reservoir, it varies as a
function of time, and the condition (\ref{PT_condition}b) just cannot be
preserved in the whole time
interval. This indicates that the features of the combined exciton-photon
system are strongly affected by the reservoir dynamics, as the inset in Fig.~%
\ref{fig_decay} shows.

In the case of exciton pumping predominantly governed by the exciton-exciton
scattering process Eq.~\eqref{RN2Chi2}, the dynamics of the system is more
complicated. The threshold value $P_{2}^{th}$ can not be derived directly
from Eq.~\eqref{eigen_without_pump}  because
of the dependence of $p_{2}$ term on the population $\left\vert \chi
\right\vert ^{2}$ of the excitonic component. Figure~\ref{fig_decay} (yellow
curve) demonstrates the behaviour of exciton and photon components below
threshold $P<P_{2}^{th}$, that was determined by numerical simulations. Although both components of the condensate are
initially amplified, then the population of $\left\vert \chi \right\vert ^{2}
$ (and hence the pumping term $p_{2}=R_{2}N^{2}\left\vert \chi \right\vert
^{2}$) attenuate very rapidly.

\begin{figure}[t]
\label{fig_decay} \includegraphics[width=0.8\linewidth]{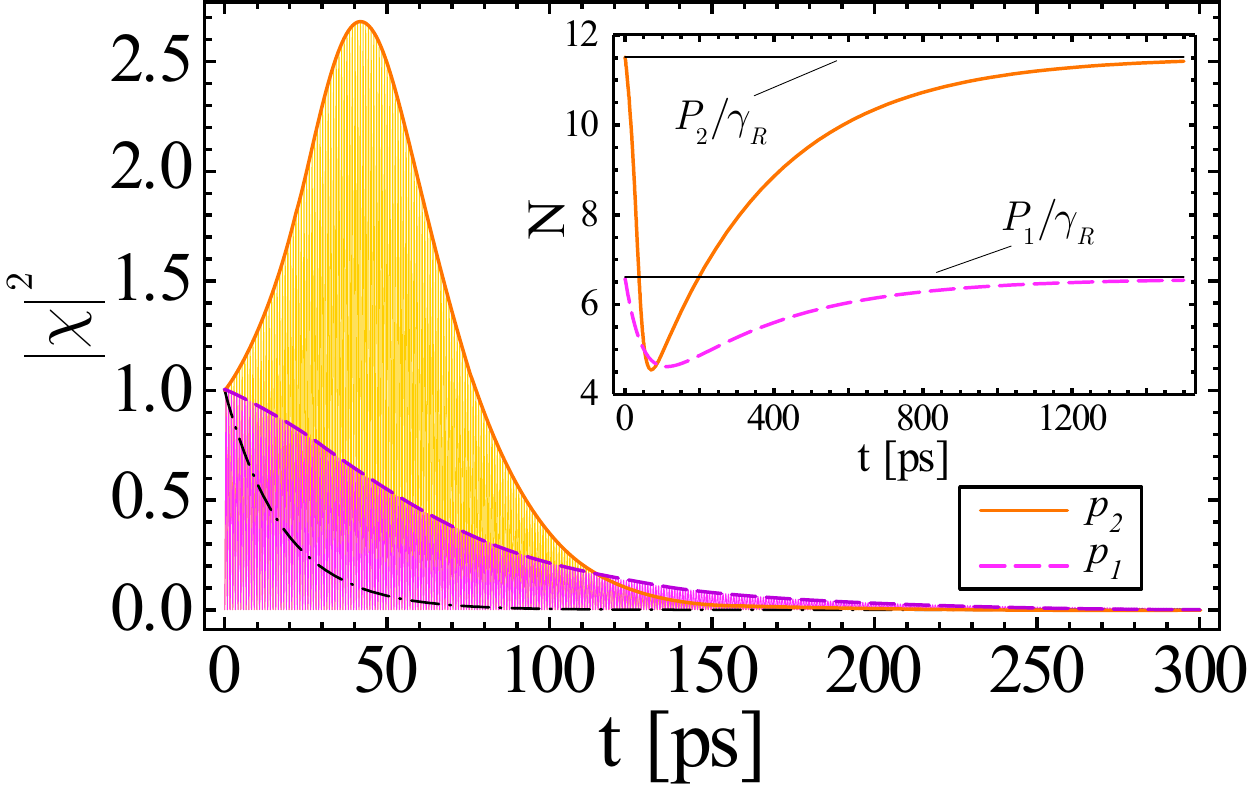}
\caption{(Color online) Time dependence of the exciton population $|\protect%
\chi |^{2} $ for $\protect\delta=0$ and for exciton state pumping $p_1$
(magenta curve) and $p_2$ (yellow curve) for cw-pump $P$ below threshold $%
P^{th}$. Dash-dotted black curve indicates decay dynamics in the absence of
reservoir. On the inset the relevant population of the reservoir $N$ for
exciton state pumping $p_1$ (magenta dashed) and $p_2$ (orange solid) is
shown. Horizontal lines correspond to the stationary reservoir
population, i.e. $N=P/\protect\gamma_R$. Parameters are: $\protect\gamma %
_{X} =0.01$~ps$^{-1} $, $\protect\gamma _{P} =0.1$~ps$^{-1} $, $\hbar \Omega
=2.5$ meV, $\protect\gamma _{R} =0.003$~ps$^{-1}$, $\hbar R_{1}
=0.01$~$\protect\mu$m$^{2}$ps$^{-1} $ and $\hbar R_{2} =0.001$~$\protect\mu$m$^{6}$%
ps$^{-1} $. Pumping rates are $P_1=0.02$~$\protect\mu$m$^{-2}$ps$^{-1} $ and
$P_2=0.035$~$\protect\mu$m$^{-2}$ps$^{-1} $ -- for magenta and yellow curves,
respectively. Initial conditions are: $\protect\chi (0)=0$, $\protect\phi %
(0)=1$, $N(0)=P/ \protect\gamma _{R} $.}
\label{fig_decay}
\end{figure}

\textit{Permanent oscillations.---}Now we analyze permanent Rabi
oscillations regime, that occurs above threshold $P>P^{th}$ and can be
preserved as long as the PT-symmetry condition is fulfilled. We represent
the solution of Eqs.(\ref{allsystem}a,b) in the form
\end{subequations}
\begin{subequations}
\label{anzats}
\begin{equation*}
\chi =\chi _{1}e^{i\omega _{1}t}+\chi _{2}e^{i\omega _{2}t},\ \ \ \phi =\phi
_{1}e^{i\omega _{1}t}+\phi _{2}e^{i\omega _{2}t}, \tag{\ref{anzats}\rm{\mbox{a,b}}}
\end{equation*}%
\noindent where $\chi _{1,2}$ and $\phi _{1,2}$ are constant amplitudes. In
the permanent oscillations regime, the reservoir population oscillates with
a small amplitude around some average value $\bar{N}=\left\langle
N\right\rangle _{t}$. Further, we assume excitonic reservoir to be in the
stationary state, $\dot{N}=0$.

Let us first consider the phonon assisted pumping of the exciton component
of the condensate. Substituting Eqs. \eqref{RN} and \eqref{anzats} into Eqs.
(\ref{allsystem}a,b) and separating real and complex parts we obtain
\end{subequations}
\begin{subequations}
\label{im_re_separated}
\begin{eqnarray}
&&R_{1}\bar{N}=\gamma _{X}+\frac{\Omega ^{2}\gamma _{P}}{\gamma
_{P}^{2}/4+\omega _{1,2}^{2}}, \\
&&\omega _{1,2}^{3}-\delta \omega _{1,2}^{2}-\omega _{1,2}(\Omega
^{2}-\gamma _{P}^{2}/4)-\delta \gamma _{P}^{2}/4=0.
\end{eqnarray}%
\noindent Clearly, only real roots of Eq. (\ref{im_re_separated}b) should be
retained for the frequencies $\omega _{1,2}$.

Let us assume that the first requirement (\ref{PT_condition}a) for the
PT-symmetry is fulfilled, i.e., we set $\delta =0$. In this case Eqs.~%
\eqref{im_re_separated} admit equal amplitudes $\left|A\right|\equiv |\chi _{1}|=|\chi
_{2}|=|\phi _{1}|=|\phi _{2}|$ and characteristic polariton frequencies
\end{subequations}
\begin{equation}
\omega _{1,2}=\pm \omega \equiv \pm \sqrt{\Omega ^{2}-\gamma _{P}^{2}/4}.
\label{freq_perm}
\end{equation}%
Equation \eqref{freq_perm} is familiar in PT-symmetry theory for coupled
oscillators,  \cite{feng11,szameit11,PRA.82403803.2010}. The value of $%
\Omega _{c}=\gamma _{P}/2$ can be associated with the PT-symmetry breaking
threshold. In the case of a driven-dissipative exciton-polariton system, we
operate significantly above the PT-threshold point, assuming that $\Omega
\gg \Omega _{c}$, due to the strong exciton-photon coupling condition.
Solution of Eq.(\ref{im_re_separated}a) with Eq.\eqref{freq_perm} leads to
the condition
\begin{equation}
R_{1}\bar{N}=\tilde{\gamma},  \label{RN_perm_condition}
\end{equation}%
that simply implies the balance between pump and loss rates in the system
and follows directly from the PT-symmetry criterion (\ref{PT_condition}b).

\begin{figure}[tbp]
\includegraphics[width=0.8\linewidth]{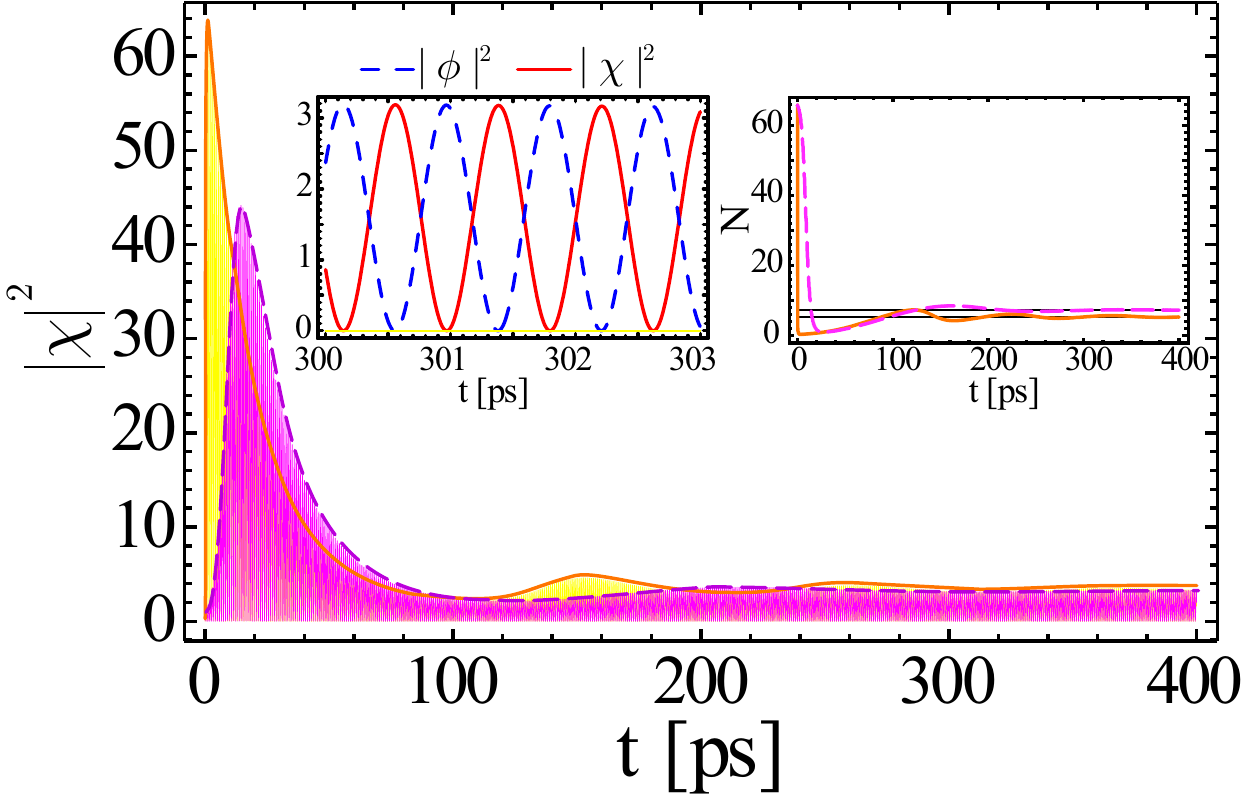}
\caption{(Color online) The same as in Fig.~\ref{fig_decay} but for cw-pump $P=0.2$~$\protect\mu$m$^{-2}$ps$^{-1} $, which is
above threshold $P^{th}$. The relevant behavior of reservoir is shown on the right panel of
the inset. Permanent Rabi oscillations for excitonic and photonic
components, plotted on a zoomed time scale, are shown on the left panel of
the inset.}
\label{fig_perm}
\end{figure}

Figure~\ref{fig_perm} demonstrates the permanent oscillations regime for
various pumping mechanisms of the excitonic component of the condensate.
Permanent oscillations, shown in the left panel of the inset in Fig.~\ref%
{fig_perm}, are established after several hundreds of picoseconds for experimentally accessible
parameters. In particular, exciton and photon populations oscillate out of
phase with equal constant amplitudes according to: $|\chi |^{2}=4|A|^{2}\cos
^{2}(\omega t)$ and $|\phi |^{2}=4|A|^{2}\sin ^{2}(\omega t)$, where $%
\left\vert A\right\vert ^{2}=\left( P-P_{1}^{th}\right) \left/ {2\tilde{%
\gamma}}\right. $ can be determined from Eqs.~(\ref{allsystem}c) and %
\eqref{RN_perm_condition}. Notably, the establishment of permanent Rabi
oscillations regime is accompanied by the quick reservoir depletion (see the
right panel on the inset in Fig.~\ref{fig_perm}) resulting in the sharp peak
of photon/exciton populations for the first few picoseconds -- see Fig.~\ref%
{fig_perm}. Undoubtedly, since the conditions \eqref{PT_condition} are
satisfied now, this regime is also capable of having dynamical PT-symmetry
properties occurring at the same time scale.

For $\delta \neq 0$ Eq.~(\ref{im_re_separated}b) possesses three real roots,
in general. Since the term $\delta \gamma _{P}^{2}/4$ might be small enough
for the moderate values of the detuning $\delta $ we can omit this term and
obtain (cf.~\eqref{freq_perm})
\begin{equation}
\omega _{1,2}=\frac{1}{2}\left( \delta \pm \sqrt{\delta ^{2}+4\Omega
^{2}-\gamma _{P}^{2}}\right) .  \label{freq_nonres}
\end{equation}%
The value of $\omega _{R}=\left\vert \omega _{1}-\omega _{2}\right\vert $
represents the frequency of Rabi oscillations.
However, since $\left\vert \omega _{1}^{{}}\right\vert \neq \left\vert
\omega _{2}^{{}}\right\vert $, Eq.~(\ref{im_re_separated}a) cannot be
satisfied for both $\omega _{1,2}$ simultaneously, cf.~\eqref{freq_perm}.
Hence, we can conclude that permanent oscillations cannot be supported in
the case of $\delta \neq 0$, which also correlates with the first PT-
symmetry criterion (\ref{PT_condition}a).

Now let us examine the permanent oscillations regime for the exciton pumping
by exciton-exciton scattering. In this case, using Eqs.~\eqref{anzats} and
omitting higher-order harmonics we obtain
\begin{equation}
R_{2}\bar{N}^{2}(|\chi _{1,2}|^{2}+2|\chi _{2,1}|^{2})=\gamma _{X}+\frac{%
\Omega ^{2}\gamma _{P}}{\gamma _{P}^{2}/4+\omega _{1,2}^{2}}.
\label{RN2_im_re_separated}
\end{equation}%
Notably Eq. (\ref{im_re_separated}b) is valid in this case as well.
Physically, it means that permanent Rabi oscillations frequency is
independent on the physical mechanism of exciton pumping. The population $%
\bar{N}$ of the excitonic reservoir can be found from the equation
\begin{equation}
P-\gamma _{R}\bar{N}-R_{2}\bar{N}^{2}\left( |\chi _{1}|^{4}+|\chi
_{2}|^{4}+4|\chi _{1}|^{2}|\chi _{2}|^{2}\right) =0.  \label{res_eq_perm}
\end{equation}%
Again, in the limit of $\delta =0$  we
obtain a simple criterion $3R_{2}\bar{N}^{2}|A|^{2}=\tilde{\gamma}$ of realization of permanent Rabi oscillations,

Figure~\ref{fig_perm} (yellow curve) demonstrates the population of
excitonic $|\chi |^{2}$ component in this case. The frequency $\omega_{R}$ of
oscillations is still determined by Eq. \eqref{freq_perm}.

Now we consider permanent Rabi-oscillations in the case of non-zero
detuning, $\delta \neq 0$, and exciton pumping $p_{2}$. Equations %
\eqref{RN2_im_re_separated} admit solutions simultaneously for both $\omega
_{1}$ and $\omega _{2}$ in the case of unequal amplitudes, $\chi _{1}\neq
\chi _{2}$. These solutions exist only for the absolute value of detuning $%
\delta $ being smaller than some critical value $\delta _{c}$. In fact, if $%
\left\vert \delta \right\vert >\left\vert \delta _{c}\right\vert $
population of an upper or a lower polariton branch (depending on the sign of
$\delta $) at the steady state becomes zero, that indicates the collapse of
Rabi oscillations. It means that the initial system (\ref{allsystem}) has a
steady state solution only; exciton $|\chi |^{2}$ and photon $|\phi |^{2}$
populations tend to some constant (non-zero) values determined by this
solution. The value of $\delta _{c}$ can be obtained using Eqs.~%
\eqref{freq_nonres} and \eqref{RN2_im_re_separated} and it approximately
reads
\begin{equation}
\delta _{c}\simeq \pm \left( \frac{\Omega ^{2}}{4\gamma _{P}}\left[ \gamma
_{X}-8\gamma _{P}+3\sqrt{\gamma _{X}^{2}+8\gamma _{P}^{2}}\right] \right)
^{1/2}.  \label{delta_crit}
\end{equation}%
It is important that $\delta _{c}$ is determined by the parameters of a
Rabi-oscillator as a whole, i.e., by the Rabi splitting frequency and
damping rates $\gamma _{X}$, $\gamma _{P}$, and it does not depend on the
parameters of the reservoir, including the external pump $P$. For GaAs-based
semiconductor microcavities we obtain $\hbar \delta _{c}\simeq 0.38\hbar \Omega
=0.96$~meV. Thus, permanent oscillations may be established in the vicinity
of exciton-photon resonance for experimentally accessible parameters.

\textit{Polarization properties.---}Here we consider a complete set of Eqs.~%
\eqref{allsystem}, that characterizes spin dependent exciton-polariton
properties. Polarization properties of the excitonic system are conveniently
described by the Stokes vector $\mathbf{S}$ with components $S_{x,y,z}$
defined by $\mathbf{S}=\frac{1}{2}\left( \Psi ^{\dag }\cdot \bm{\sigma}\cdot \Psi
\right)$,
where $\Psi =(\chi _{+},\chi _{-})^{\mathrm{T}}$ and $\sigma _{x,y,z}$ are
the Pauli matrices. Since $\delta _{+}\neq \delta _{-}$ the behavior of
polarization of exciton component is completely determined by detuning $%
\Delta $ for a given non-zero Zeeman shift $\Delta _{Z}\neq 0$.

\begin{figure}[tbp]
\includegraphics[height=0.33\linewidth]{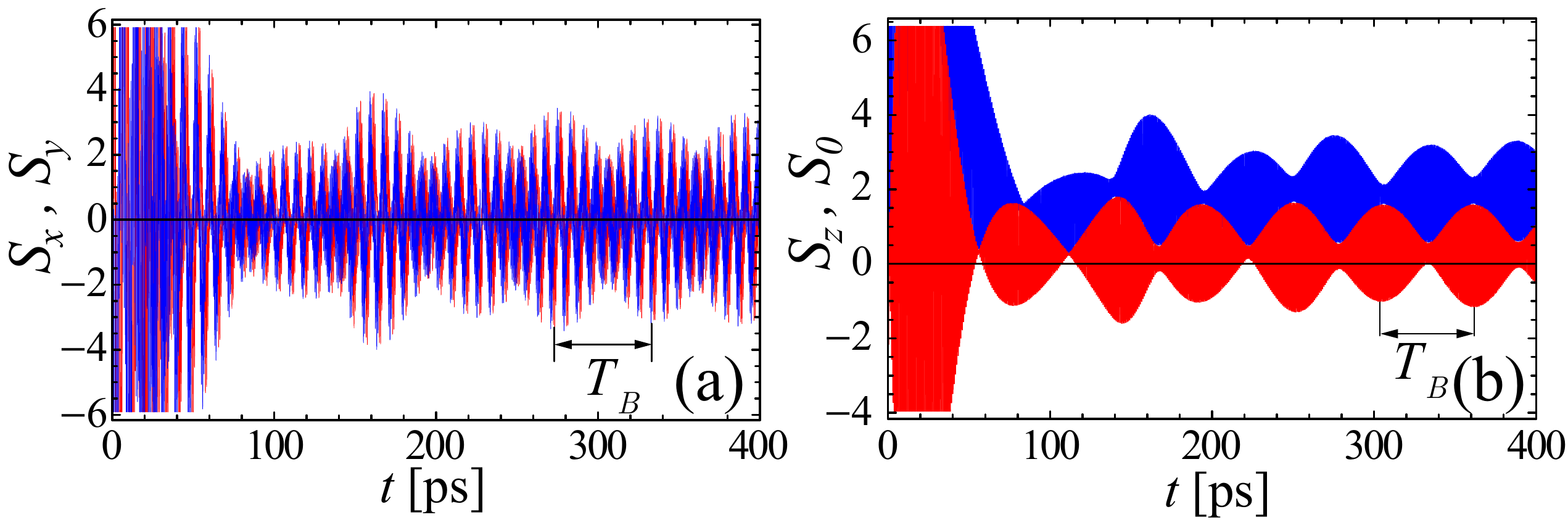} %
\includegraphics[width=0.45\linewidth]{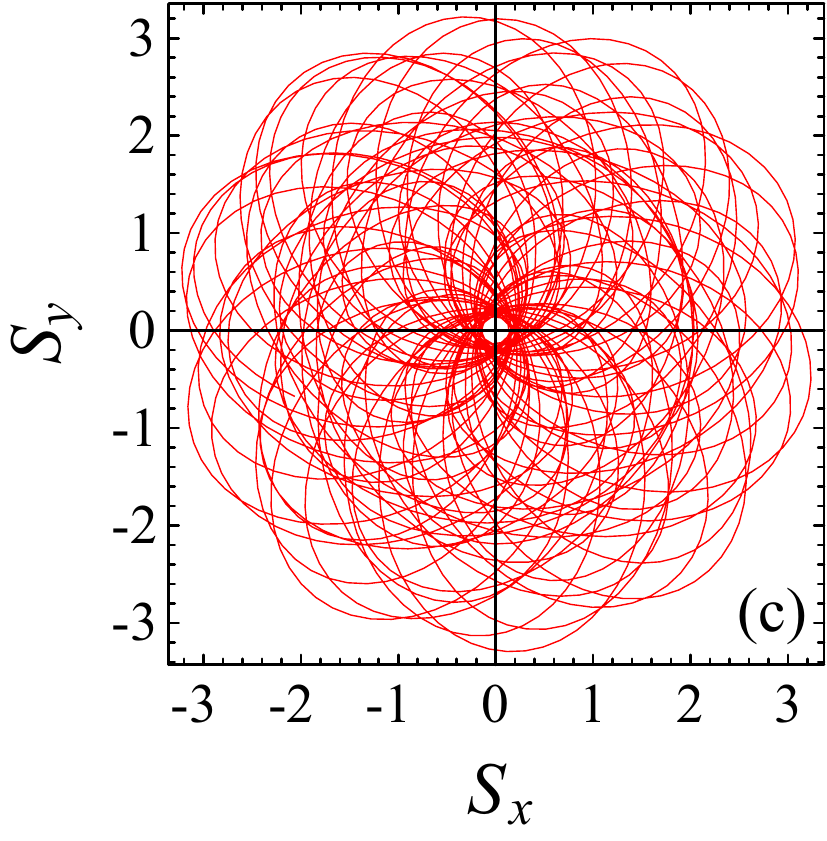}\ \ \ \ \ \ \ \ %
\includegraphics[width=0.4\linewidth]{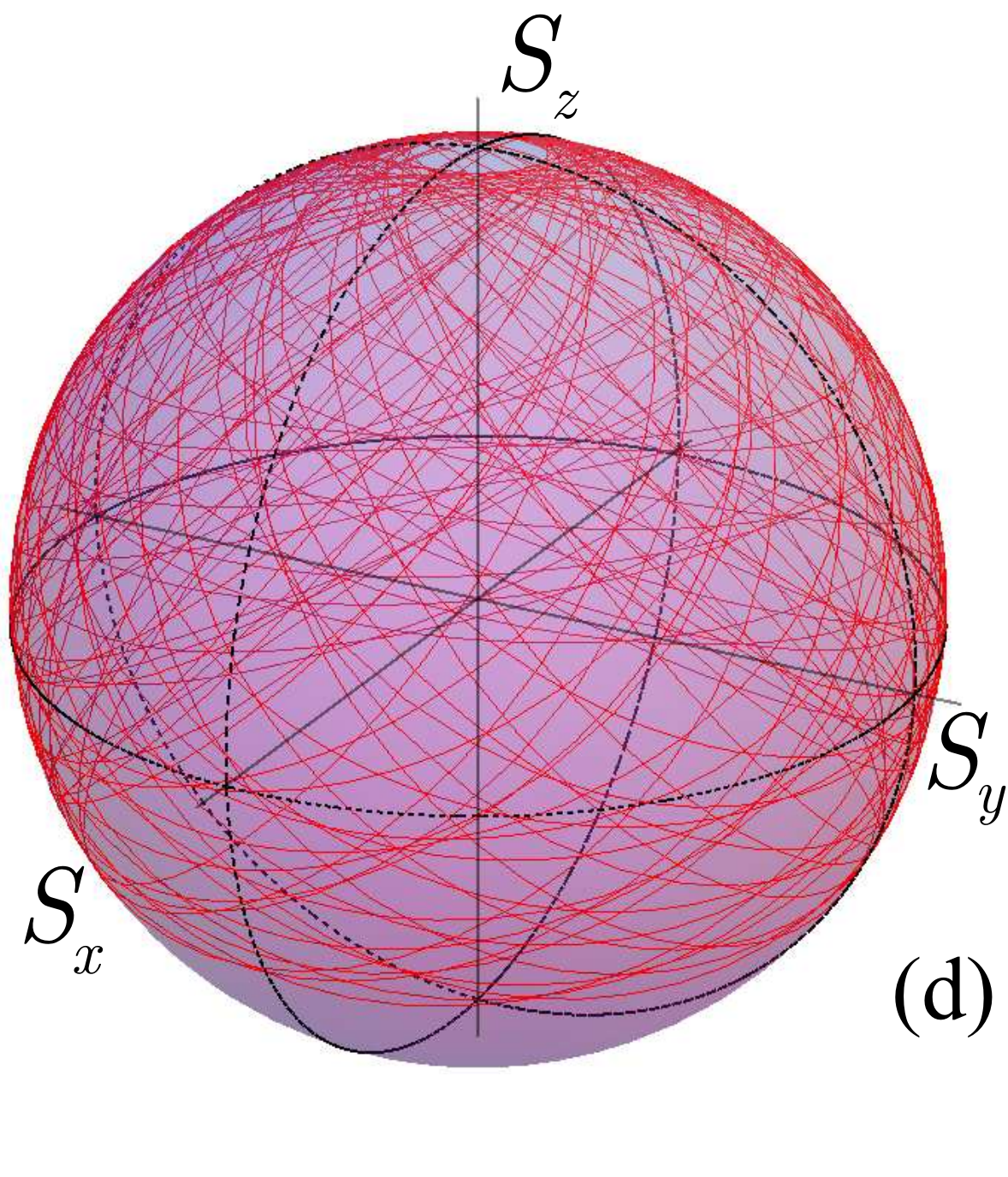}
\caption{Time evolution of Stokes parameters for $\Delta_Z=0.2\Omega$, $%
\Delta=0.15\Omega$. $S_x$ (red) and $S_y$ (blue) components -- (a); $S_z$
(red) and $S_0$ (blue) component -- (b); (c) and (d) show the evolution of
polarization on $\left(S_x,S_y\right)$-plane (c) and on Stokes sphere (d) on
the time scale of $T_B=2%
\protect\pi/\protect\omega_B$ characterizing permanent Rabi oscillations
beating period. The parameters are: $p_X[N]=p_2$, $P=0.2$~$\protect\mu$m$%
^{-2}$ps$^{-1}$; other values are the same as in Fig.~\protect\ref{fig_decay}.
}
\label{fig.polarization2}
\end{figure}

\begin{figure}[tbp]
\includegraphics[width=0.47\linewidth]{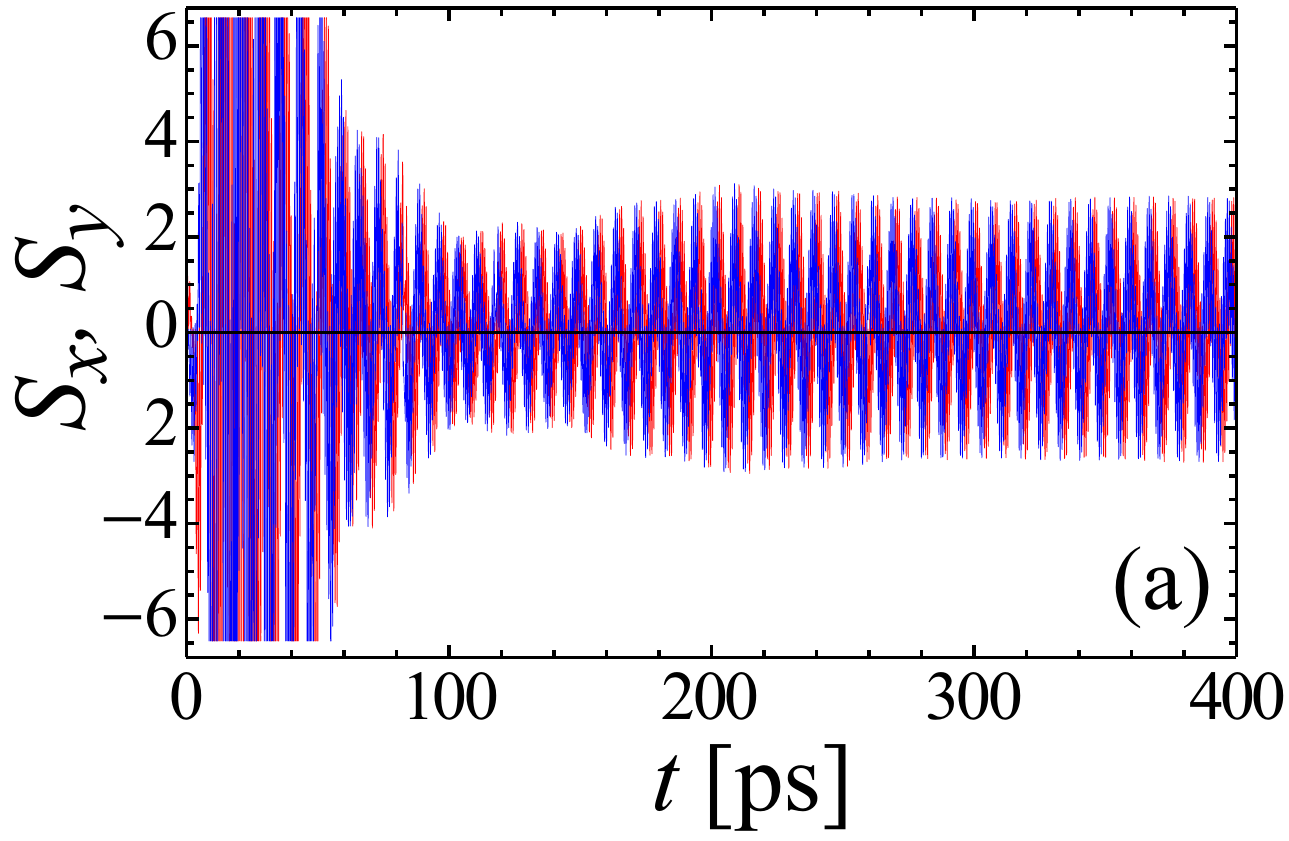} \includegraphics[width=0.47%
\linewidth]{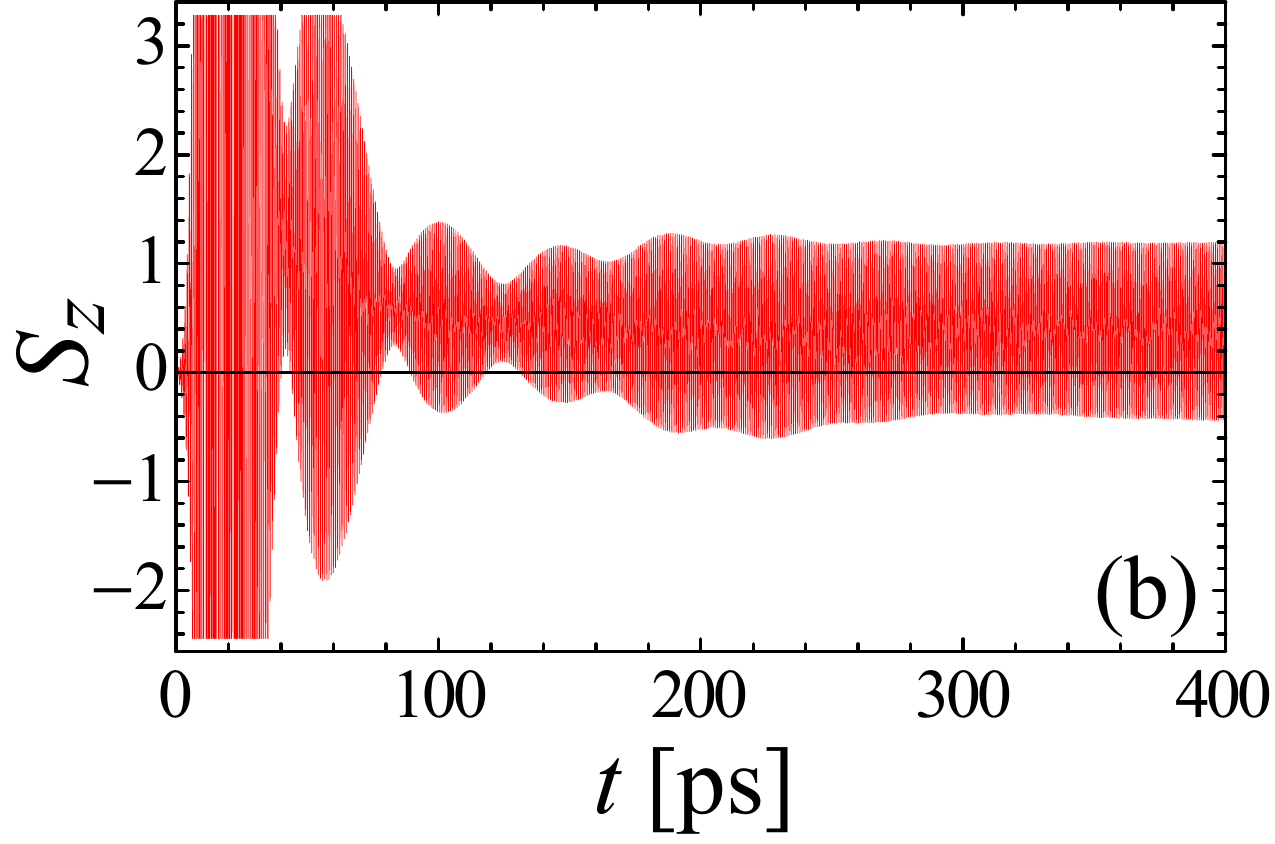}
\caption{Time evolution of Stokes parameters $S_{x,y,z}$ for $%
\Delta_Z=\Delta=0.2\Omega$. The parameters are: $p_X[N]=p_1$, $P=0.2$~$%
\protect\mu$m$^{-2}$ps$^{-1}$; other values are the same as in Fig.~\protect
\ref{fig_decay}}
\label{fig.polarization3}
\end{figure}

The general case corresponding to both $\Delta $ and $\Delta _{Z}$ are
nonzero is illustrated by Fig.~\ref{fig.polarization2}. Using
anzats \eqref{anzats} and definition for Stokes vector $\mathbf{S}$ it is easy to
show that components $S_{x}$ and $S_{y}$ oscillate with the
frequency $\Delta _{Z}$ (frequent beats in Figs.~\ref{fig.polarization2}a and \ref{fig.polarization3}a). Besides there are fast oscillations with the frequency $\left(\omega _{R+}+\omega _{R-} \right)/2$ close to the frequency of Rabi-oscillations.  If the conditions of
establishing of permanent oscillations are fulfilled for both spin
components simultaneously ($\left\vert \delta _{\pm }\right\vert <\delta _{c}
$), the frequencies of Rabi oscillations $\omega _{R+}$ and $\omega _{R-}$
are different, because of the relation $\left\vert \delta _{+}\right\vert
\neq \left\vert \delta _{-}\right\vert $. In this case, the amplitudes of
Stokes vector oscillations undergo additional long-period beats with the
period $T_{B}=2\pi /\omega _{B}=2\pi /\left( \omega _{R+}-\omega _{R-}\right)
\approx \frac{2\pi \Omega }{\Delta \Delta _{Z}}$, see Fig.~\ref%
{fig.polarization2}(b). These beats are sustained in time.

In Fig.~\ref{fig.polarization2}(d) we represented the evolution of the
Stokes vector $\mathbf{S}$ normalised to the total number of excitons $S_{0}=%
\frac{1}{2}\left( \left\vert \chi _{+}\right\vert ^{2}+\left\vert \chi
_{-}\right\vert ^{2}\right) $ on the Poincare sphere with a unit radius.
Note that $S_{0}$, describing the excitonic subsystem, oscillates in time as
shown in Fig.~\ref{fig.polarization2}(b).

Let us consider the limit where permanent oscillations disappear for one of the spin
components, i.e. for $\left\vert \delta _{+}\right\vert <\left\vert \delta
_{c}\right\vert $ and $\left\vert \delta _{-}\right\vert >\left\vert \delta
_{c}\right\vert $, or vice versa, that can be achieved by the appropriate
choice of $\Delta _{Z}$ and $\Delta $. In fact, in this case the beats in
Stokes parameters with the period $T_{B}$ are suppressed in time and the
amplitude of oscillations tends to the constant value. Figure~\ref%
{fig.polarization3} demonstrates these features for Stokes parameters in the
specific case of $\Delta =\Delta _{Z}.$ In this limit, the detunings $\delta
_{-}=0$ and $\delta _{+}=2\Delta _{Z}$. Physically, it means that permanent
oscillations can be established for an exciton pumping mechanism described
by \eqref{RN}, but only for \textquotedblleft $-$\textquotedblright\ spin
component while population of another component will approach a constant value. At long time intervals only fast oscillations survive.

\textit{In conclusion}, we have described the regime of permanent Rabi
oscillations in a driven-dissipative exciton-photon system feeded from the
incoherent exciton reservoir. Two types of reservoir-Rabi-oscillator
coupling mechanisms have been examined. We have shown that the permanent
Rabi oscillation regime is established above the threshold pumping where the
dynamical PT-symmetry of the coupled exciton-photon system is realised.
Notably, this regime occurs in a non-resonant case only if the exciton
component of the condensate is pumped by exciton-exciton scattering
predominantly. We studied excitonic polarization properties in the presence
of external magnetic fields. The conditions for permanent Rabi oscillations
for polarized excitonic and/or photonic system are established and
discussed. Realization of permanent Rabi oscillations in this case play a
crucial role for creation and manipulating of long-lived spin polarisation
in the exciton-photon system which pave the way to the design of new optical
memory devices.

\begin{acknowledgments}
\textit{Acknowledgments.---}A.P.A. acknowledges useful discussions with Yuri Kivshar. This work was supported by RFBR Grants No.
14-02-31443, No. 14-02-92604, No. 14-02-97503, No. 15-52-52001, No.
15-59-30406, No. 14-32-50438 and EU project PIRSESGA-2013-612600 LIMACONA.
The financial support from the Russian Ministry of Education and Science
(Contract No.11.G34.31.0067 with SPbSU and leading scientist A. V. K.) is
acknowledged. A.P.A. acknowledges support from ``Dynasty''\ Foundation and
Y.G.R. acknowledges support from CONACYT under the grant No.\ 251808.
\end{acknowledgments}

\end{document}